\documentclass[twocolumn,superscriptaddress,floatfix,preprintnumbers,prd ,nofootinbib,hyperref]{revtex4-2}
\usepackage[colorlinks=true,breaklinks=true]{hyperref}
\usepackage[utf8]{inputenc}
\hypersetup{allcolors=[rgb]{0.0 0.0 0.6},linkcolor=[rgb]{0.0 0.0 0.6}}
\usepackage{mathtools}
\usepackage[dvipsnames]{xcolor}
\usepackage{float}
\usepackage{siunitx}
\usepackage{letltxmacro}
\usepackage{enumitem}
\usepackage[normalem]{ulem}
\LetLtxMacro{\oldcite}{\cite}
\renewcommand{\cite}[1]{\mbox{\oldcite{#1}}}
\usepackage{orcidlink}
\long\def\exclude#1{}
\allowdisplaybreaks
\bibpunct{[}{]}{,}{n}{}{,}
\definecolor{color1}{rgb}{0.8,0.0,0.0}
\definecolor{color2}{rgb}{0.0,0.8,0.0}
\definecolor{color3}{rgb}{0.0,0.0,0.8}
\definecolor{color4}{rgb}{0.5,0.0,0.5}

\begin{document}
\preprint{TIFR/TH/25-7, N3AS-25-004}

\title{Stronger Constraints on Primordial Black Holes as Dark Matter Derived from the \\Thermal Evolution of the Intergalactic Medium over the Last Twelve Billion Years}

\author{Nabendu Kumar Khan\,\orcidlink{0009-0005-7052-3012}}
\email{nabendu.khan@tifr.res.in}
\affiliation{Tata Institute of Fundamental Research, Homi Bhabha Road, Mumbai 400005, India}

\author{Anupam Ray\,\orcidlink{0000-0001-8223-8239}}
\email{anupam.ray@berkeley.edu}
\affiliation{Department of Physics, University of California, Berkeley, California 94720, USA}

\author{Girish Kulkarni\,\orcidlink{0000-0001-5829-4716}}
\email{kulkarni@theory.tifr.res.in}
\affiliation{Tata Institute of Fundamental Research, Homi Bhabha Road, Mumbai 400005, India}

\author{Basudeb Dasgupta\,\orcidlink{0000-0001-6714-0014}}
\email{bdasgupta@theory.tifr.res.in}
\affiliation{Tata Institute of Fundamental Research, Homi Bhabha Road, Mumbai 400005, India}

\begin{abstract}
    Primordial black holes (PBHs) have been explored as potential dark matter candidates, with various astrophysical observations placing upper limits on the fraction $f_\mathrm{PBH}$ of dark matter in the form of PBHs. However, a largely underutilized probe of PBH abundance is the temperature of the intergalactic medium (IGM), inferred from the thermal broadening of absorption lines in the \mbox{Lyman-$\alpha$} forest of quasar spectra. PBHs inject energy into the IGM via Hawking radiation, altering its thermal evolution.  In this work, we constrain this energy injection by self-consistently modeling its interplay with the cosmological ultraviolet background from galaxies and supermassive black holes. Leveraging IGM temperature measurements spanning the past twelve billion years ($z \sim 0$ to $6$), we derive one of the most stringent constraints on PBH-induced heating from light PBHs within the mass range $10^{15}$--$10^{17}$~g. Specifically, for $M_\mathrm{PBH} = 10^{16}$~g, we find $f_\mathrm{PBH} < 5 \times 10^{-5}$ at 95\% confidence, with the bound scaling approximately as $M_\mathrm{PBH}^{4}$ at other masses.  Our inclusion of helium reionization and low-redshift temperature measurements strengthens previous IGM-based PBH constraints by an order of magnitude or more. Compared to other existing limits, our result is among the strongest, second only to the constraints from the 511 keV line from the Galactic Center, but with distinct systematics. More broadly, this study highlights the IGM thermal history as a powerful and independent probe of beyond-standard-model physics.
\end{abstract}

\maketitle
\section{Introduction}
The intergalactic medium (IGM) is the dominant reservoir of matter in the Universe.  As such, the thermal evolution of the IGM is a potentially promising probe of any process that affects the thermal state of the Universe. 

It is now known that the IGM at redshifts \( z < 6 \) is in thermal equilibrium with a cosmological radiation background created by star-forming galaxies and accreting supermassive black holes. 
The balance between photoheating from this radiative background, heating from gravitational collapse, and cooling from Hubble expansion imparts to the IGM a power-law temperature–density relation, usually written as \cite{1997MNRAS.292...27H},
\begin{equation}
    T = T_{\mathrm{IGM}} \Delta^{\gamma - 1},
\end{equation}
where \( T_{\mathrm{IGM}} \) is the temperature at mean density, \( \Delta = \rho / \bar{\rho} \) represents the overdensity, and \( \gamma-1 \) is the power-law index. The parameters \( T_{\mathrm{IGM}} \) and \( \gamma \) evolve with redshift, encoding information about the evolution of the IGM, the radiation background influencing it, and the associated heating and cooling processes.

Fortunately, the temperature of the IGM can also be measured out at least to $z\sim 6$.  Most of these measurements use observations of the Lyman-\(\alpha\) forest to infer the IGM temperature. The Lyman-$\alpha$ forest is a series of absorption lines observed in the spectra of distant luminous quasars and serves as an excellent tracer of the intervening diffuse gas (see Refs.~\cite{2009RvMP...81.1405M, 2016ARA&A..54..313M} for reviews). The individual absorption lines in the Lyman-$\alpha$ forest are caused by the neutral hydrogen residing in the IGM. Consequently, the widths of these lines are sensitive to the thermal state of the IGM. The thermal motion of neutral hydrogen introduces a broadening in the width of the lines, expressed as $\Delta \nu = \nu_{\mathrm{Ly}\alpha} b_\mathrm{th}/c$, where $c$ is the speed of light, $\nu_{\mathrm{Ly}\alpha}$ is the Lyman-$\alpha$ absorption frequency, and $b_\mathrm{th} = \sqrt{2k_\mathrm{B}T/m_\mathrm{H}}$ is the Doppler parameter associated with the thermal motion with $k_\mathrm{B}$ being the Boltzmann's constant, $m_\mathrm{H}$ the mass of the hydrogen atom, and $T$ the temperature of the IGM. By directly measuring the Doppler parameter of individual absorption lines in the Lyman-$\alpha$ forest, or by indirectly measuring it from other statistics sensitive to thermal broadening, it is therefore possible to infer the temperature of the IGM \cite{2000MNRAS.318..817S, 2000ApJ...534...41R, 2001ApJ...557..519Z, 2002ApJ...567L.103T, 2010ApJ...718..199L, 2011MNRAS.410.1096B, 2015ApJ...799..196L, 2017PhLB..773..258G, 2017Sci...356..418R, 2017MNRAS.466.2690R, 2018MNRAS.474.2871R, 2019ApJ...872..101B, 2019ApJ...872...13W, 2019ApJ...876...71H, 2019ApJ...876...31G, 2021MNRAS.506.4389G, 2020MNRAS.494.5091G, 2022ApJ...933...59V, 2025MNRAS.536....1H}. This approach has revealed that $T_\mathrm{IGM}$ is of the order of $\sim 10^4$~K in the twelve billion years since $z\sim 6$, with some evolution. At $z\sim 6$, the temperature is around $11,500$~K at $z \sim 5.8$ and gradually declines to $\sim 7,300$~K by $z\sim4.6$. Due to helium reionization at $z\sim3$, the temperature peaks at approximately $14,750$~K before cooling again to $\sim 4,700$~K at $z \sim 0.06$. The typical uncertainties in these measurements range from $10\%$ to $20\%$.

Exotic energy injection processes can significantly affect the thermal and ionization history of the IGM. Examples include dark matter decay or annihilation{~\cite{Slatyer:2015jla,Slatyer:2015kla,Liu:2016cnk}}, dark matter-baryon interactions{~\cite{Ali-Haimoud:2015pwa,Munoz:2015bca}}, ultralight dark photon dark matter{~\cite{Mirizzi:2009iz,McDermott:2019lch}}, and light primordial black holes (PBHs){~\cite{Clark:2016nst,Mittal:2021egv}}. Nonetheless, measurements of the IGM temperature from the Lyman-$\alpha$ forest as a probe of these processes have largely been overlooked so far. Indeed, while the Lyman-$\alpha$ forest has been used to constrain exotic processes, these constraints have primarily relied on measurements of the Lyman-$\alpha$ transmission power spectrum, which probes small-scale cosmological matter density fluctuations; these constraints have largely disregarded the information contained in the Lyman-$\alpha$ forest about the thermal state of the IGM. Examples include warm dark matter \cite{2005PhRvD..71f3534V, 2008PhRvL.100d1304V, 2013PhRvD..88d3502V, 2017PhRvD..96b3522I, 2017JCAP...12..013B, 2020JCAP...04..038P, 2023PhRvD.108b3502V, 2023arXiv230904533I}, sterile neutrinos \cite{2005PhRvD..71f3534V, 2008PhRvL.100d1304V, 2017JCAP...12..013B, 2019MNRAS.489.3456G}, fuzzy dark matter \cite{2017PhRvL.119c1302I, 2017MNRAS.471.4606A}, ultralight axions \cite{2021PhRvL.126g1302R}, primordial black holes (PBHs) with masses around $100$~M$_{\odot}$ \cite{2019PhRvL.123g1102M}, and isocurvature fluctuations \cite{2020PhRvD.101l3518I}. The potential of IGM temperature as a diagnostic tool for exotic physics remains largely untapped \cite{2017JCAP...11..043M, 2021PhRvD.104d3514L, 2022PhRvL.129u1102B, 2024arXiv240910617S}.

In this study, we revisit the use of IGM temperature measurements from the Lyman-$\alpha$ forest to place constraints on light PBHs. Such light PBHs in the mass range of $10^{15}$ to $10^{17}\,\mathrm{g}$  have recently gained attention as potential dark matter candidates. Through Hawking evaporation, they emit particles that can be detected by various ground-based and space-based instruments. Non-detection of these Hawking emitted particles has been used to place significant constraints on their abundance \cite{Carr:2009jm, Boudaud:2018hqb, Arbey:2019vqx, DeRocco:2019fjq, Laha:2019ssq, Dasgupta:2019cae, Laha:2020ivk, Chan:2020vsr, Ray:2021mxu, Coogan:2020tuf, Kim:2020ngi, Laha:2020vhg, Chan:2020zry, Ray:2022rfz, Bernal:2022swt, 2024JCAP...09..022T, 2024PhRvD.110l3022L} (see also recent reviews on this topic~\cite{Carr:2020gox,Carr:2020xqk,Green:2020jor, Escriva:2022duf}). Beyond direct detection, the energy deposited into the IGM by Hawking radiation also leaves indirect imprints that can be probed through cosmological observations, including the cosmic microwave background (CMB) \cite{Clark:2016nst, Stocker:2018avm, Poulter:2019ooo, Acharya:2020jbv, Cang:2020aoo} and the global \mbox{21-cm} signal \cite{Clark:2018ghm, Mittal:2021egv, Natwariya:2021xki, Cang:2021owu, Saha:2021pqf}. Recently, Ref.~\cite{2024arXiv240910617S} derived constraints on the abundance of light PBHs using IGM temperature measurements from the Lyman-$\alpha$ forest. In this paper, we significantly improve and extend their argument (see Section~\ref{section5}) — arriving at an order of magnitude stronger constraint, which is more reliable and includes more detailed microphysics.

\section{Hawking Radiation from Evaporating Primordial Black Holes}
\begin{figure}
    \centering
    \includegraphics*[width=\columnwidth]{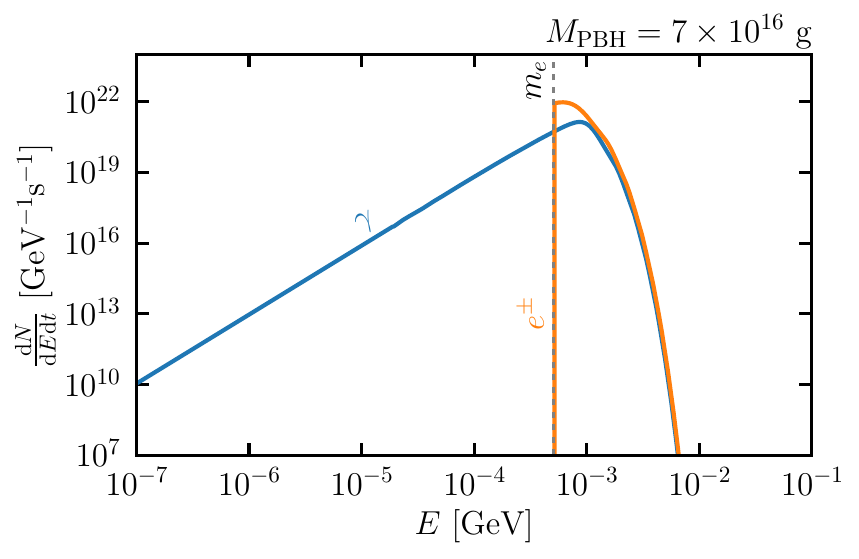}
    \caption{The energy spectrum of photons, electrons and positrons emitted by a PBH with mass \( M_\mathrm{PBH} = 7 \times 10^{16} \, \mathrm{g} \). The spectra peak at \( E \sim 1 \, \mathrm{MeV} \) and display a blackbody-like profile: an exponential decline at higher energies and a power-law trend at lower energies. The electron spectrum features a kinematic cutoff at the electron rest mass.}
    \label{fig1}
\end{figure}

Non-rotating and uncharged PBHs of mass $M_\mathrm{PBH}$ have a temperature that depends solely on their mass~\cite{Page:1976df,Page:1976ki,MacGibbon:1990zk,MacGibbon:2007yq}
\begin{equation} \label{equ1}
    k_\mathrm{B}T_\mathrm{PBH} = \frac{\hbar c^3}{8 \pi G_\mathrm{N} M_\mathrm{PBH}} = 1 \, \mathrm{GeV} \left( \frac{10^{13} \, \mathrm{g}}{M_\mathrm{PBH}} \right),
\end{equation}
where  \( \hbar = h / 2\pi \) is the reduced Planck's constant, \( G_\mathrm{N} \) denotes the gravitational constant. These PBHs can emit Standard Model (SM) particles, such as photons, electrons, positrons etc, via Hawking radiation, gradually shedding their mass. The spectrum of these emitted particles can be parametrized as~\cite{Page:1976df,Page:1976ki,MacGibbon:1990zk,MacGibbon:2007yq}
\begin{equation} \label{equ2}
    \frac{\mathrm{d} N}{\mathrm{d}E \, \mathrm{d}t} = \frac{1}{2\pi \hbar
} \frac{\Gamma (E, \mu, T_\mathrm{PBH})}{\exp\left({\frac{E}{k_\mathrm{B} T_\mathrm{PBH}}}\right) - (-1)^{2s}},
\end{equation}
where \( k_\mathrm{B} \) is the Boltzmann constant, \( E \) is the total energy of the emitted particles, \( s \) is the spin of the emitted particle, and \( \mu \) denotes its rest mass. \( \Gamma(E, \mu, T_\mathrm{PBH}) \) denotes the dimensionless gray-body factor, representing the probability of an emitted particle escaping the black hole's potential, and is evaluated numerically. We use the publicly available code \texttt{BlackHawk} \cite{2019EPJC...79..693A} to calculate the \textit{primary} Hawking emission spectra for various particles, and in Figure~\ref{fig1}, we illustrate the emitted electron, positron, and photon spectra for a PBH of mass \( 7 \times 10^{16} \, \mathrm{g} \). We do not show the neutrino spectrum, as the effect of neutrinos on the IGM is negligible. We also neglect potential \textit{secondary} contributions in order to be conservative.

These Hawking-emitted particles from light PBHs can interact with the IGM, depositing energy into the gas. The rate of this energy injection can be expressed as  
\begin{align}
    \left( \frac{\mathrm{d} E}{\mathrm{d} V \, \mathrm{d} t} \right)^\mathrm{inj} &= 
    2 n_\mathrm{PBH} \int \left( \frac{\mathrm{d} N}{\mathrm{d} E_e \, \mathrm{d} t} \right)_{e^{\pm}}  
    (E_e - m_e c^2) \, \mathrm{d}E_e \nonumber \\
    &\quad +  n_\mathrm{PBH} \int \left( \frac{\mathrm{d} N}{\mathrm{d} E_\gamma \, \mathrm{d} t} \right)_\gamma  
     E_\gamma \, \mathrm{d}E_\gamma.
     \label{eqn:energy_injection}
\end{align}
where $n_\mathrm{PBH}$ denotes the number density of light PBHs. Assuming a monochromatic mass distribution, this number density can be expressed as 
\begin{equation} 
    n_\mathrm{PBH} = f_\mathrm{PBH} \frac{\rho_c \Omega_\mathrm{DM}}{M_\mathrm{PBH}} (1 + z)^3, \label{equ4}
\end{equation}
where \( f_\mathrm{PBH} \) is the fraction of dark matter composed of PBHs, \( \rho_c \) is the critical density, and \( \Omega_\mathrm{DM} \) is the dark matter density parameter~\cite{Aghanim:2018eyx}.

\section{IGM Thermal and Ionization History in Presence of PBH Energy Injection}
Injected high-energy particles cool rapidly via various processes \cite{2009PhRvD..80d3526S}. Electrons and positrons cool through inverse Compton scattering with the CMB and atomic processes like collisional ionization and excitation of hydrogen and helium atoms, and Coulomb heating of free electrons and ions in the IGM, while photons cool through redshifting, pair production, photon-photon scattering, Compton scattering, and photoionization of the hydrogen and helium atoms in the IGM. Considering the timescales of these processes, the injected energy from the PBHs can be assumed to deposit into four channels: (\textit{a}\/) hydrogen ionization, (\textit{b}\/) helium ionization, (\textit{c}\/) Lyman-\( \alpha \) excitation, and (\textit{d}\/) Coulomb heating. 
Following Ref.~\cite{2020PhRvD.101b3530L}, we encapsulate this in a numerical factor \( f_\mathrm{Cnl}(z, \textbf{x}) \), allowing the energy deposition rate into channel `Cnl' to be expressed as  
\begin{equation}
    \left( \frac{\mathrm{d} E}{\mathrm{d} V \, \mathrm{d} t} \right)^\mathrm{dep}_\mathrm{Cnl} = f_\mathrm{Cnl}(z, \textbf{x}) \left( \frac{\mathrm{d} E}{\mathrm{d} V \, \mathrm{d} t} \right)^\mathrm{inj}. \label{f_c}
\end{equation}
The deposition fractions in the four channels, \textit{viz.}, \( f_\mathrm{H,ion} \), \( f_\mathrm{He,ion} \), \( f_\mathrm{exc} \), and \( f_\mathrm{heat} \), correspond to energy fractions allocated to hydrogen ionization, helium ionization, Lyman-\( \alpha \) excitation, and Coulomb heating, respectively. These factors, \( f_\mathrm{Cnl}(z, \textbf{x}) \), depend on both the redshift \( z \) and the ionization state of the IGM, represented by \( \textbf{x} = (x_\mathrm{HII}, x_\mathrm{HeII}, x_\mathrm{HeIII}) \), where the ionization fractions are defined as \( x_\mathrm{HII} = n_\mathrm{HII} / n_\mathrm{H} \), \( x_\mathrm{HeII} = n_\mathrm{HeII} / n_\mathrm{H} \), and \( x_\mathrm{HeIII} = n_\mathrm{HeIII} / n_\mathrm{H} \), with \( n_\mathrm{H} \) being the total hydrogen number density, \( n_\mathrm{HII} \) being the number density of singly ionized hydrogen, \( n_\mathrm{HeII} \) the number density of singly ionized helium, and \( n_\mathrm{HeIII} \) the number density of doubly ionized helium. Since the ionization state determines the abundance of free electrons, it directly affects the efficiency of various energy deposition processes. Consequently, \( f_\mathrm{Cnl}(z, \textbf{x}) \) values are redshift-dependent and largely influenced by the reionization process. 

\begin{figure}
    \centering
    \includegraphics*[width=\columnwidth]{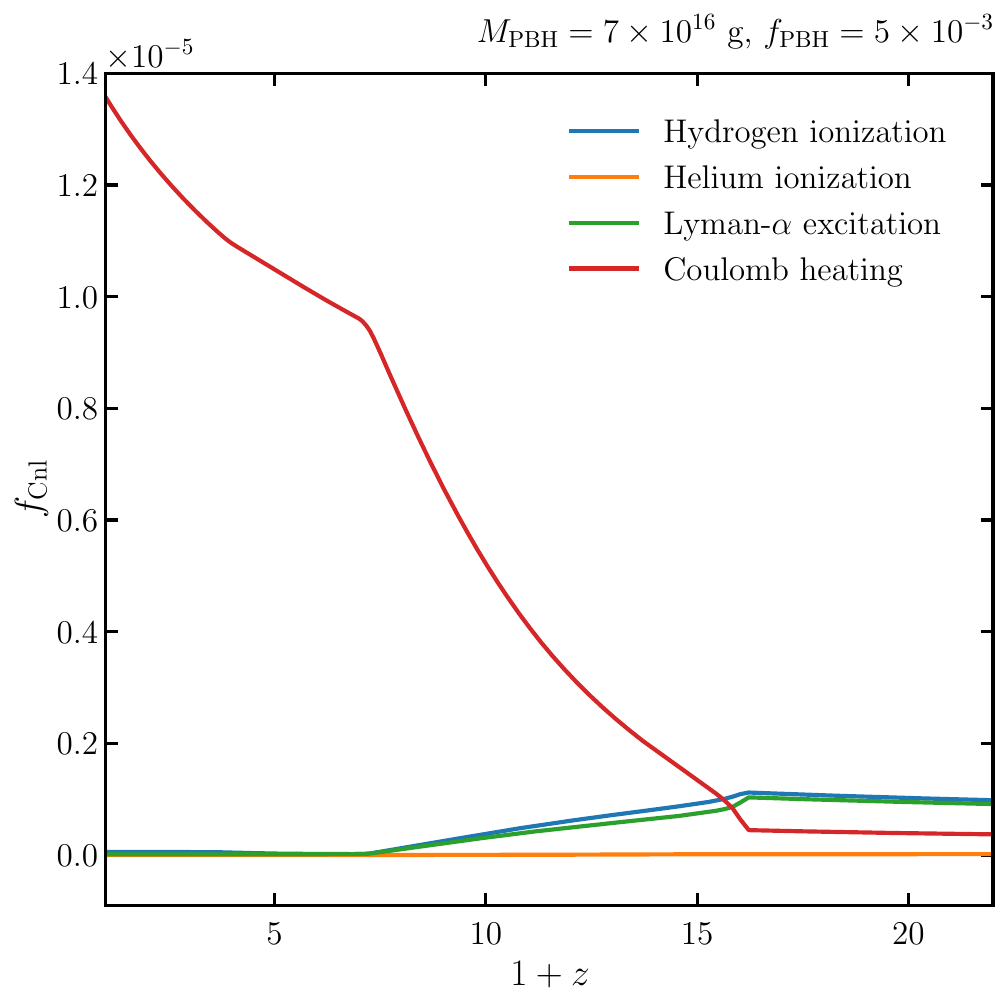}
    \caption{Values of the quantity $f_\mathrm{Cnl}$, which describe the fraction of the energy injected by PBHs that is deposited into the IGM through channel `Cnl', as a function of redshift. These fractions are found self-consistently via a computation of the thermochemical evolution of the IGM in the presence of a UV background produced by galaxies and supermassive black holes.  For this figure, we fix $M_{\text{PBH}} = 7 \times 10^{16}$ and $f_{\text{PBH}} = 5 \times 10^{-3}$.  The channels shown are hydrogen ionization, helium ionization, Lyman-$\alpha$ excitation, and Coulomb heating.  We see a strong contribution of PBHs to the IGM heating, and negligible contribution to IGM ionization.}
    \label{fig2}
\end{figure}

Figure~\ref{fig2} shows the redshift evolution of the calculated $f_\mathrm{Cnl}$ values for all four channels mentioned above with $M_{\text{PBH}} = 7 \times 10^{16}$\,g and $f_{\text{PBH}} = 5 \times 10^{-3}$.  During and after reionization, the increased abundance of free charged particles enhances interactions with the injected high-energy electrons and positrons. As a result, Coulomb heating becomes more efficient, significantly affecting the thermal evolution of the IGM. On the other hand, the ionization fraction of the IGM remains largely unaffected, as there is no significant change in the $f$ values for ionization. This is because the injected particles from PBHs have much higher energies, and the cross section for ionization is comparatively small.  The fraction of injected energy deposited as heat is represented by \( f_{\text{heat}}(z, \textbf{x}) \), given by
\begin{equation}
    f_\mathrm{heat}(z, \textbf{x}) = \frac{2 n_\mathrm{PBH}}{\left(\frac{\mathrm{d}E}{\mathrm{d} V  \mathrm{d}t}\right)^\mathrm{inj}} \int \left( \frac{\mathrm{d}N}{\mathrm{d} E \mathrm{d}t}\right)^\mathrm{inj}_{e^\pm}  R_\mathrm{heat} (E) \, \mathrm{d} E, \label{equ8}
\end{equation} 
where \( R_\mathrm{heat}(E) \) represents the total energy deposited into Coulomb heating, accounting for the cumulative energy transfer from an injected electron of energy \( E \) until its energy is fully dissipated. The function \( R_\mathrm{heat}(E) \) is obtained by solving the recursive equation given in Equation~\ref{S5}. A detailed derivation of \( R_\mathrm{heat}(E) \) is provided in the Appendix~\ref{appendix2}. The heating efficiency, characterized by \( f_{\text{heat}}(z, \textbf{x}) \), evolves with redshift, and depends on the ionization state of the IGM. At lower redshifts, the increasing abundance of free electrons enhances the energy deposition rate, making Coulomb scattering highly efficient. As a result, a significant fraction of the injected energy is rapidly converted into heat, leading to a rise in the IGM temperature. This strong PBH-induced heating effect plays a crucial role in shaping the thermal history of the IGM, thereby strengthening constraints on PBH abundance.

With the values of \( f_\mathrm{Cnl} \) determined, we can now calculate the energy deposition rate into the IGM due to PBHs. This energy deposition rate is then used to evolve the temperature and ionization fractions of the IGM. The temperature evolution equation is given by
\begin{equation}
    \dot{T}_\mathrm{IGM} = \dot{T}_\mathrm{adia} + \dot{T}_\mathrm{C}  + \dot{T}_\mathrm{atom} + \dot{T}_\mathrm{reion}  + \dot{T}_\mathrm{PBH}, \label{temp}
\end{equation} 
where \( \dot{T}_\mathrm{adia} \) represents the cooling due to adiabatic expansion, and \( \dot{T}_\mathrm{C} \) accounts for heating or cooling due to (inverse) Compton scattering off the CMB. The term \( \dot{T}_\mathrm{atom} \) encompasses all atomic cooling processes such as collisional ionization, collisional excitation, recombination, and Bremsstrahlung cooling. (Details on these processes are provided in Appendix~\ref{appendix1}.) The term \( \dot{T}_\mathrm{reion} \) includes the effect of photoheating during reionization, driven by active galactic nuclei (AGNs) and star-forming galaxies. To compute $\dot{T}_\mathrm{reion}$, we adopt the reionization model of Puchwein \textit{et al.}~\cite{2019MNRAS.485...47P}, which self-consistently accounts for the evolution of galaxies and AGN, the resulting ultraviolet background (UVB), and its impact on the IGM by solving the radiative transfer equation. This approach yields a reionization and thermal history that is consistent with observational constraints as well as three-dimensional radiative transfer simulations of reionization.  Puchwein \textit{et al.}\ also demonstrate that applying their photoheating and photoionization rates to a gas cell at mean density, as we do in this work, accurately reproduces the temperature and ionization evolution predicted by full cosmological simulations (see their Figure C1).  Additionally, we apply a uniform rescaling factor of 0.9 to the H$\,$\textsc{i}, He$\,$\textsc{i}, and He$\,$\textsc{ii} photoheating rates in the Puchwein \textit{et al.}\ model to better match the observed IGM temperature. As the spectral shape of the ionizing background remains uncertain in UVB models, such rescaling is physically motivated: changing the photoheating rates is effectively equivalent to modifying the hardness of the ionizing spectrum (e.g., see Table 4 of \cite{2021MNRAS.506.4389G}). Finally, $\dot{T}_\mathrm{PBH}$, denotes the contribution of PBH energy injection to the temperature evolution of the IGM, and is given by  
\begin{equation}
    \dot{T}_\mathrm{PBH} = \frac{2}{3 (1 + \mathcal{F}_{\text{He}} + x_e) n_\text{H}} \left(\frac{\mathrm{d}E}{\mathrm{d}V \, \mathrm{d}t}\right)^{\mathrm{dep}}_{\mathrm{heat}}, \label{equ7}
\end{equation}  
where \( \mathcal{F}_\mathrm{He} = n_\mathrm{He} / n_\mathrm{H} \) is the helium-to-hydrogen number density ratio, for which we adopt \( \mathcal{F}_\mathrm{He} = 0.08 \) in this work. 

Next, the evolution of the hydrogen and singly ionized helium ionization fraction \( x_\mathrm{HII/HeII} \) is described by
\begin{equation}
    \dot{x}_{\text{HII/HeII}} = \dot{x}_{\text{HII/HeII}}^{\mathrm{atom}}  + \dot{x}_{\text{HII/HeII}}^{\text{reion}} + \dot{x}_{\text{HII/HeII}}^{\text{PBH}}, \label{ion}
\end{equation} 
where \( \dot{x}_{\text{HII/HeII}}^{\mathrm{atom}} \) accounts for collisional ionization and recombination (see Appendix~\ref{appendix1} for details), and \( \dot{x}_{\text{HII/HeII}}^{\text{reion}} \) corresponds to the photoionization rates from the reionization model \cite{2019MNRAS.485...47P}.  The contribution from PBHs to hydrogen ionization, is given by
\begin{equation}
    \dot{x}_{\text{HII}}^{\text{PBH}} = \frac{1}{\mathcal{R} n_\text{H}} \left(\frac{\mathrm{d}E}{\mathrm{d}V \, \mathrm{d}t}\right)^{\text{dep}}_{\mathrm{H,ion}} + \frac{(1 - \mathcal{C})}{0.75 \mathcal{R} n_\text{H}} \left(\frac{\mathrm{d}E}{\mathrm{d}V \, \mathrm{d}t}\right)^{\text{dep}}_{\mathrm{exc}},
\end{equation}
where \( \mathcal{R} = 13.6 \, \mathrm{eV} \) is the hydrogen ionization potential, and \( \mathcal{C} \) is the Peebles-C factor, which represents the probability that a hydrogen atom in the \( n = 2 \) state decays to the ground state before photoionization can occur. PBH-induced ionization and Lyman-$\alpha$ excitation arise from both injected electrons/positrons and photons. Electrons and positrons contribute through collisional ionization and excitation, while photons induce photoionization and photoexcitation. The fractions of the total injected energy deposited into ionization and excitation channels are denoted by \( f_{\text{H,ion}}(z, \textbf{x}) \) and \( f_{\text{exc}}(z, \textbf{x}) \), respectively. These energy deposition processes directly impact the ionization history of the IGM, altering its thermal and ionization evolution.

The PBH-induced ionization of helium is given by  
\begin{equation}
    \dot{x}_{\text{HeII}}^{\text{PBH}} = \frac{1}{\mathcal{R}_{\text{He}} n_\text{He}} \left(\frac{\mathrm{d}E}{\mathrm{d}V \, \mathrm{d}t}\right)^{\text{dep}}_{\mathrm{He,ion}}, 
\end{equation}  
where \( \mathcal{R}_{\text{He}} = 24.6 \, \mathrm{eV} \) is the ionization potential of helium.

For doubly ionized helium, the evolution equation is given by
\begin{equation} \dot{x}_{\text{HeIII}} = \dot{x}_{\text{HeIII}}^{\mathrm{atom}} + \dot{x}_{\text{HeIII}}^{\text{reion}}, \label{ion_HeIII} \end{equation}
where the atomic processes, including collisional ionization and recombination, contribute as
\begin{equation} \dot{x}_{\mathrm{HeIII}}^{\mathrm{atom}} = x_{\mathrm{HeII}} n_e \Gamma_{e \mathrm{HeII}}^{\mathrm{ion}} - x_{\mathrm{HeIII}} n_e \alpha_{A,\mathrm{HeIII}},\end{equation}
with collisional ionization and recombination coefficients taken from Ref.~\cite{2007MNRAS.374..493B}. The contribution from reionization is
\begin{equation} \dot{x}_{\mathrm{HeIII}}^{\mathrm{reion}} = x_{\mathrm{HeII}} \Gamma_{\gamma \mathrm{HeII}}^{\mathrm{ion}}. \end{equation}
The photoionization rates used in this calculation are taken from Ref.~\cite{2019MNRAS.485...47P}.

To compute the IGM thermal history in the presence of PBHs, we now need to solve the equations for $f_\mathrm{Cnl}$, $T_\mathrm{IGM}$, $x_\mathrm{HII}$, $x_\mathrm{HeII}$, and $x_\mathrm{HeIII}$ self-consistently. We do this using the publicly available code \texttt{DarkHistory} \cite{2020PhRvD.101b3530L}, by modifying the code to solve these equations down to $z=0$, and incorporating He$\,$\textsc{iii} evolution as well as PBH energy injection.  We now describe these modifications to \texttt{DarkHistory}.

\begin{figure*}
    \centering
    \includegraphics*[width=0.95\textwidth]{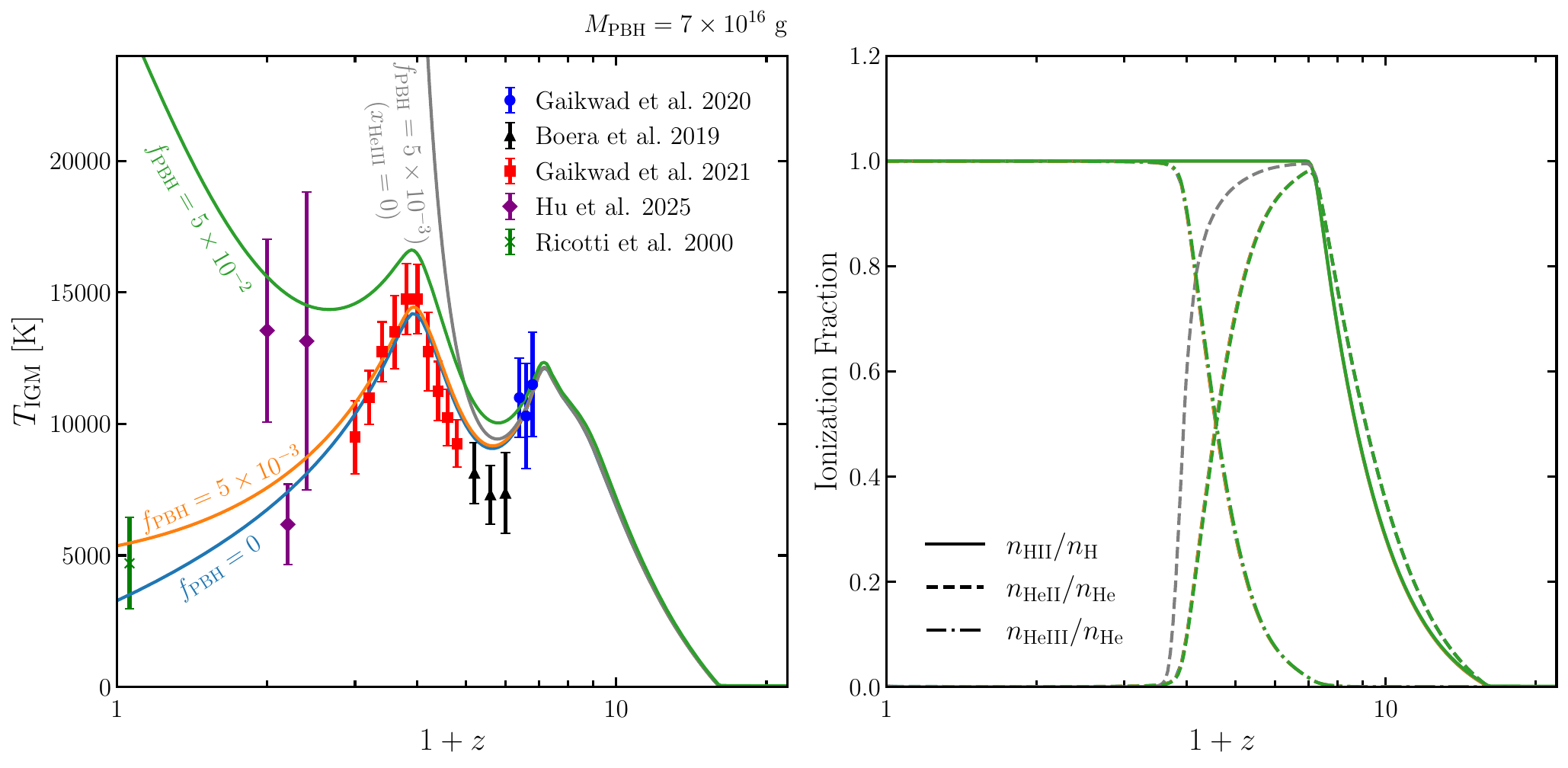}
    \caption{Left panel shows the evolution of temperature of the IGM at mean density with redshift for $M_\mathrm{PBH} = 7 \times 10^{16}$\,g, with three $f_\mathrm{PBH}$ values, 0 (no PBHs; blue curve), $5 \times 10^{-3}$ (orange curve), and $5 \times 10^{-2}$ (green curve), in comparison with various measurements from $z\sim 0$--$6$. The right panel shows the evolution of the ionization fractions of hydrogen (solid curve), singly-ionized helium (dashed curve), and doubly-ionized helium (dot-dashed curve).  These ionization fractions remain largely unaffected by the inclusion of PBHs; it is the IGM temperature that shows the more significant effect. The gray curves in both panels represent a model without He$\,$\textsc{iii} evolution for \( f_\mathrm{PBH} = 5 \times 10^{-3} \), which leads to an unphysical temperature increase.
    }
    \label{fig3}
\end{figure*}

\section{Modifications to \NoCaseChange{\texttt{DarkHistory}}}
In order to solve our thermochemistry, we have modified the publicly available code \texttt{DarkHistory} \cite{2020PhRvD.101b3530L}.  We make three modifications to this code: (\textit{a}\/) addition of PBH energy injection, (\textit{b}\/) inclusion of He$\,$\textsc{iii} evolution, and (\textit{c}\/) extension of the calculations to $z=0$.  We now describe these modifications.  In order to incorporate energy injection by PBHs, we include the electron and positron spectra from Equation~(\ref{equ2}) a single PBH into the \texttt{in\_spec\_elec} data structure and the photon spectrum, again following from Equation~(\ref{equ2}), into the \texttt{in\_spec\_phot} data structure in the \texttt{main.evolve} function. Additionally, we modify the \texttt{phys.inj\_rate} function to account for the total energy injection rate from PBHs according to Equation~(\ref{eqn:energy_injection}).  To include He$\,$\textsc{iii} evolution, we modify the function \texttt{tla.get\_history}, which solves the temperature and ionization evolution equations at each time step $\Delta t$.  Instead of setting $x_{\mathrm{HeIII}}$ to zero, as is the case in the public version of \texttt{DarkHistory}, we now self-consistently use the values resulting from an equation of the helium ionization equations at each time step.  Finally, to extend the calculations to $z=0$, we modify the \texttt{main.evolve} function. Incorporating He$\,$\textsc{iii} chemistry and achieving calculations down to $z=0$ also necessitates recalculating the deposition fractions $f_\mathrm{Cnl}$. In the public version of \texttt{DarkHistory}, values of $f_\mathrm{Cnl}$ are precomputed for a range of redshifts and ionization fractions, without considering He$\,$\textsc{iii} chemistry and only extending down to $z \sim 3$. To overcome these constraints, we adapt \texttt{DarkHistory} to calculate $f_\mathrm{Cnl}$ dynamically and self-consistently at each time step. This is accomplished by integrating the deposition fraction computation directly into the \texttt{main.evolve} function, simultaneously incorporating He$\,$\textsc{iii} chemistry. This enhancement in calculating deposition fractions represents the central methodological improvement in this paper compared to previous works. We initialise \texttt{DarkHistory} at $z=3,000$ using default initial values for all quantities.

Although the original \texttt{DarkHistory} code was developed for high-redshift applications ($z \gtrsim 3$) and primarily models hydrogen and singly ionized helium, our modified version extends its validity down to $z = 0$ by including He\,\textsc{iii} evolution and adopting the reionization model of Puchwein \textit{et al.}\ \cite{2019MNRAS.485...47P}. This extension allows a self-consistent treatment of late-time photoheating and photoionization. Since PBH evaporation acts as a decay process, with energy injection scaling linearly with PBH density, it is not enhanced by clustering or halo formation, making our constraints robust to clustering uncertainties. However, we note that spatial inhomogeneities and detailed astrophysical feedback from galaxies and AGNs are not explicitly modeled. The use of mean-density photoheating rates ensures realistic average thermal histories, but small-scale variations and local reionization effects remain beyond the scope of this study. Future work could incorporate these aspects through full radiative-transfer simulations.

\section{Results from Modified \NoCaseChange{\texttt{DarkHistory}}} \label{section5}
Using the coupled Equations~\ref{temp}, \ref{ion} and \ref{ion_HeIII}, we calculate the temperature and ionization histories as functions of redshift for a range of $M_\mathrm{PBH}$ and $f_\mathrm{PBH}$ values. These calculations begin at redshift \( 1 + z = 3000 \) with initial conditions \( x_\mathrm{HII} = 1 \) and \( x_\mathrm{HeII} = 1 \). The model-predicted temperatures are then compared with observational data, which will be described in detail in the next section, for the redshift range $0.06 < z < 5.8$. Figure~\ref{fig3} shows the temperature evolution calculated for $M_\mathrm{PBH} = 7 \times 10^{16}$\,g, along with the dataset. We examine $f_\mathrm{PBH}$ values of 0, $5 \times 10^{-3}$, and $5 \times 10^{-2}$. A value of 0 indicates the absence of PBHs, and this baseline model fits the data very well. For the other two $f_\mathrm{PBH}$ values, the temperature increases at lower redshifts due to enhanced Coulomb heating (as seen in Figure~\ref{fig2}), with this effect being most prominent at the lowest redshifts. By visual inspection, we can exclude the $f_\mathrm{PBH} = 5 \times 10^{-2}$ model, as it significantly overpredicts the IGM temperature and does not provide a good fit to the data. The evolution of the ionization fraction is shown in the right panel of Figure~\ref{fig3}, where the ionization fraction remains unchanged after the inclusion of PBHs, with differences always below 1\%. This is expected, as the $f_\mathrm{ion}$ values for ionization are close to zero, even for large $f_\mathrm{PBH}$ values such as $5 \times 10^{-2}$. This suggests that PBH-induced energy injection predominantly affects the thermal rather than the ionization properties of the IGM.

Figure~\ref{fig3} also demonstrates how our analysis improves upon Ref.~\cite{2024arXiv240910617S}, enabling more stringent constraints on PBH dark matter from IGM temperature measurements. A critical distinction between our approach and Ref.~\cite{2024arXiv240910617S} is their assumption that $x_\mathrm{HeIII}=0$. However, neglecting He$\,$\textsc{iii} evolution impacts the ionization state of the IGM, \textbf{x}, and the free electron number density, $n_e$. This, in turn, affects the energy deposition efficiency, $f_\mathrm{heat}(z, \textbf{x})$ (Equation~\ref{equ8}), and consequently modifies the heating rate, $\dot{T}_\mathrm{PBH}$ (Equation~\ref{equ7}). Additionally, omitting He$\,$\textsc{iii} alters both photoionization and photoheating rates, resulting in an unphysical temperature evolution.

To explicitly evaluate the impact of neglecting He$\,$\textsc{iii}, we performed a comparative calculation by artificially setting $x_\mathrm{HeIII}=0$ and contrasting it with our comprehensive model, which fully incorporates He$\,$\textsc{iii} evolution. As illustrated in Figure~\ref{fig3}, the gray curve corresponds to the model without He$\,$\textsc{iii}, while the orange curve represents our full calculation for $M_\mathrm{PBH}=7\times10^{16}$g and $f_\mathrm{PBH}=5\times10^{-3}$. This comparison clearly demonstrates that neglecting He$\,$\textsc{iii} initiates a significant deviation in temperature at, in this case, $z \lesssim 7$ (left panel), which intensifies toward lower redshifts. Furthermore, omitting He$\,$\textsc{iii} evolution noticeably impacts the He$\,$\textsc{ii} fraction beginning around $z \sim 7$ (right panel). Intriguingly, Ref.~\cite{2024arXiv240910617S} also reported a similar, unexpected temperature increase, visible as a teal curve in Figure~1 of their paper. Our findings strongly suggest that this temperature anomaly results directly from neglecting He$\,$\textsc{iii} evolution, emphasizing the necessity of including comprehensive helium evolution to accurately constrain PBH abundance.

Furthermore, the analysis in Ref.~\cite{2024arXiv240910617S} disregards IGM temperature data at $z < 3.5$, thereby significantly weakening their constraints. As Figure~\ref{fig3} demonstrates, cumulative heat injection from PBHs becomes particularly substantial at lower redshifts, highlighting the importance of extending the analysis to $z=0$.

A third critical distinction pertains to the treatment of reionization and photoheating. Ref.~\cite{2024arXiv240910617S} employs a simplified reionization model that does not self-consistently address IGM thermochemistry or incorporate current understandings of hydrogen reionization and its subsequent impacts. In contrast, we adopt the physically motivated reionization model by Puchwein \textit{et al.}\ \cite{2019MNRAS.485...47P}, which explicitly models ionizing radiation from known populations of active galactic nuclei (AGN) and star-forming galaxies, together with a realistic density distribution of the inhomogeneous IGM. The Puchwein \textit{et al.}\ model also compared favorably with more sophisticate three-dimensional radiative transfer cosmological simulations of the IGM that are known to reproduce several IGM observations.  Conversely, Ref.~\cite{2024arXiv240910617S} merely adjust their reionization history to align with Planck measurements without explicitly considering the astrophysical sources. Their ``photoheated'' model simplifies pre-reionization heating as $\dot{T}\star = \dot{x}_\mathrm{HII}(1+\chi)\Delta T$, where $\Delta T$ is an empirical parameter ranging between $2\times10^4$~K and $3\times10^4$~K, and post-reionization heating relies solely on the spectral index of the ionizing background. While flexible, this approach fails to explicitly track contributions from distinct astrophysical mechanisms.

In summary, our analysis differs crucially from Ref.~\cite{2024arXiv240910617S} by explicitly accounting for He$\,$\textsc{iii} evolution, extending the analysis down to redshift zero, and adopting a physically comprehensive reionization model. Overall, on the one hand, these improvements reduce the temperature enhancement due to PBHs.  On the other hand, these changes allow us to use measurements from a much wider redshift range than before, which leads directly to significantly tighter constraints on the abundance of PBH dark matter, as will be discussed further below.

\section{The Temperature Measurements of the Intergalactic Medium from the Lyman-$\alpha$ Forest}
We now describe the IGM temperature measurements presented in Figure~\ref{fig3} and used below to derive constraints on PBHs. Gaikwad \textit{et al.}\ 2020~\cite{2020MNRAS.494.5091G} investigated high-resolution Lyman-$\alpha$ spectra from QSOs at $6 < z \lesssim 6.5$ to study the incidence of Lyman-$\alpha$ transmission spikes. By comparing the incidence of Lyman-$\alpha$ transmission spikes with IGM simulations, they found that the width and height of the transmission spikes are sensitive to the IGM's temperature. This lead to an inference of $T_\mathrm{IGM} = (11,000 \pm 1600, 10,500 \pm 2100, 12,000 \pm 2200)$\,K at $z = 5.4, 5.6, 5.8$ respectively.  Boera \textit{et al.}\ 2019~\cite{2019ApJ...872..101B} use a power-law model for the IGM temperature density relation to measure the thermal state of the IGM at $4 < z < 5.2$ using 1D Lyman-$\alpha$ transmission power spectrum data from 15 quasar spectra obtained from Keck/HIRES and VLT/UVES. They also introduced a new parameter, $u_0$, to account for pressure smoothing, representing the cumulative energy per proton deposited into a gas parcel at the mean background density. From their analysis, they inferred a temperature $T_\mathrm{IGM} \sim 7000$--$8000$\,K with no significant evolution with redshift.  In a separate study, Gaikwad \textit{et al.}\ 2021~\cite{2021MNRAS.506.4389G} focused on the thermal state of the IGM in the redshift range $2 \leq z \leq 4$. They analyzed 103 high-resolution Lyman-$\alpha$ spectra using four transmission distribution statistics: the transmission power spectrum, Doppler parameter distribution, wavelet statistics, and curvature statistics. By comparing these measurements with cosmological hydrodynamic simulations, they inferred a temperature of $T_\mathrm{IGM} = 14,750 \pm 1,322$ K at $z \sim 3$, which they attributed to He$\,$\textsc{ii} reionization. Recently, Hu \textit{et al.}\ 2025~\cite{2025MNRAS.536....1H} extended IGM temperature measurements to lower redshifts, providing the first direct constraints on the thermal and ionization state of the IGM at $0.9 < z < 1.5$ by modeling the distribution of line widths and neutral hydrogen column density values of individual absorption lines in the Lyman-$\alpha$ forest. They analyzed 301 Lyman-$\alpha$ absorption lines from 12 archival HST spectra.  Their measurements yielded the following IGM temperatures at mean density: \( T_\mathrm{IGM} = 13,520 \, \mathrm{K} \) at \( z = 1.4 \), \( T_\mathrm{IGM} = 6,020 \, \mathrm{K} \) at \( z = 1.2 \), and \( T_\mathrm{IGM} = 13,080 \, \mathrm{K} \) at \( z = 1\). 
Ricotti \textit{et al.}\ 2000~\cite{2000ApJ...534...41R} measured the temperature parameter at very low redshift using the line width and column density distribution. They analyzed 79 lines from 15 Seyfert galaxies and quasars detected in observations with the HST/GHRS. Their results yielded a temperature of $T_\mathrm{IGM}=4,700$ K at $z=0.06$, with an error of $1,739$ K.  All of these data are shown in the left panel of Figure~\ref{fig3}.  We now use these measurements to derive constraints on PBHs.
\section{Constraints on \NoCaseChange{PBHs} from IGM temperature}
To constrain PBH parameters from temperature measurements, we perform a \(\chi^2\) difference test. The \(\chi^2\) statistic is defined as\begin{equation}
    \chi^2 = \sum_{i=1}^{20} \frac{\left( T_{\mathrm{IGM}, i}^{\mathrm{data}} - T_{\mathrm{IGM}, i}^{\mathrm{model}} \right)^2}{\sigma_i^2},
\end{equation}  
where \( T_{\mathrm{IGM}, i}^{\mathrm{data}} \) is the observed temperature at redshift \( z_i \), \( T_{\mathrm{IGM}, i}^{\mathrm{model}} \) is the predicted temperature from a given model, and \( \sigma_i \) is the reported measurement uncertainty for that redshift bin. For the no-PBH scenario, the calculated $\chi^2$ value is 18.0, which is well within the acceptable range, indicating a good fit to the data. To assess whether the inclusion of PBHs significantly worsens the fit to the data, we compute the difference in \(\chi^2\) between the PBH model and the no-PBH model:
\begin{equation}
    \Delta \chi^2 = \chi^2_{\mathrm{PBH}} - \chi^2_{\textrm{no-PBH}}.
\end{equation}
The PBH model introduces two additional free parameters: \( f_\mathrm{PBH} \) (the PBH fraction) and \( M_\mathrm{PBH} \) (the PBH mass), which increase the degrees of freedom by 2. According to the $\chi^2$ distribution with two degrees of freedom, the threshold value for the \(\chi^2\) difference test at the 95\% confidence level is approximately 5.99.

\begin{figure}
    \centering
    \includegraphics*[width=0.5\textwidth]{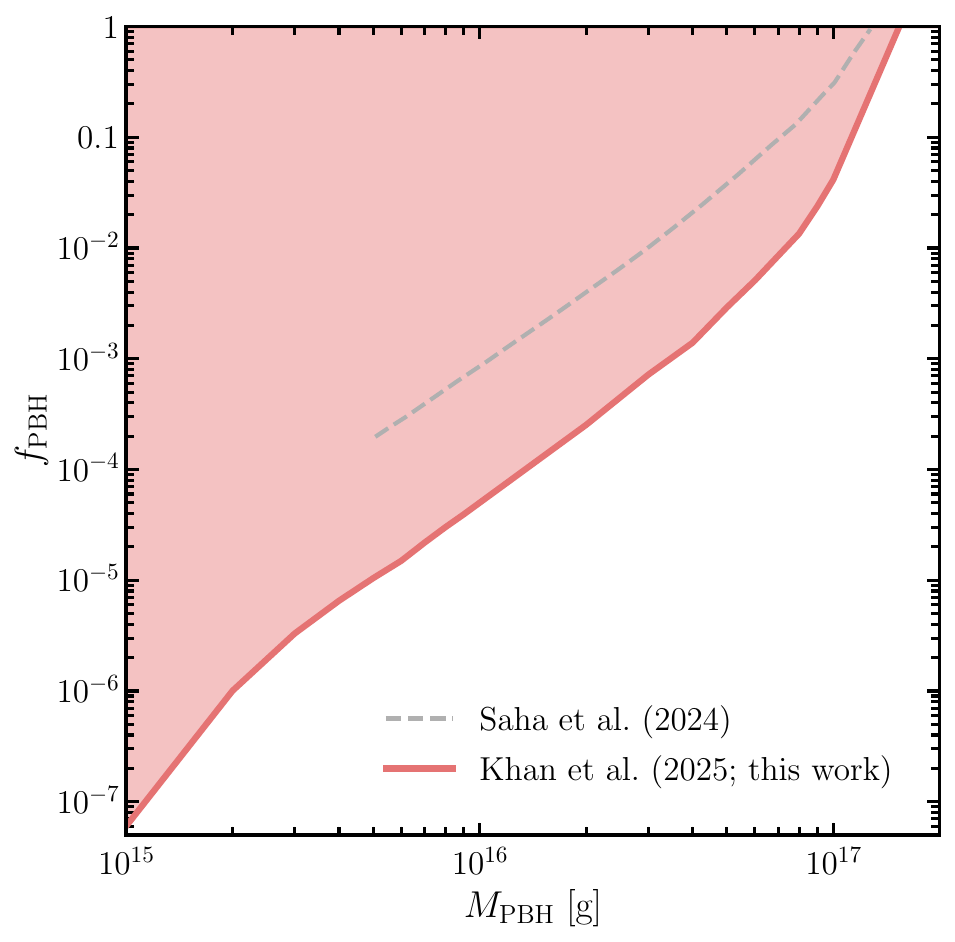}
    \caption{Our limits on \( f_\mathrm{PBH} \) for PBH masses between \( 10^{15} \,\mathrm{g} \) and \( 2\times 10^{17} \,\mathrm{g} \), inferred at the 95\% confidence level from measurements of the IGM temperature between redshifts 0 and 6, are shown in the figure. The shaded region represents the excluded parameter space based on observational data. For comparison, we also show similar constraints from Ref.~\cite{2024arXiv240910617S}, depicted in gray, where we have adopted their strongest reported constraint from the ``photoheating" model variations. Our constraints are stronger than those from Ref.~\cite{2024arXiv240910617S} by an order of magnitude or more.}
    \label{fig4}
\end{figure}

The resulting constraints on PBH parameters are shown in Figure~\ref{fig4}. The red curve represents the upper limit on $f_\mathrm{PBH}$ as a function of PBH mass, $M_\mathrm{PBH}$. The red-shaded region corresponds to the excluded parameter space at the 95\% confidence level. At $M_\mathrm{PBH} = 7 \times 10^{16}$\,g, we constrain $f_\mathrm{PBH} \leq 8.6 \times 10^{-3}$. The constraints weaken for larger PBH masses due to the slower evaporation rate of more massive PBHs. Since Hawking radiation causes PBHs to lose mass at a rate $\dot{M}_\mathrm{PBH} \propto M_\mathrm{PBH}^{-2}$ \cite{1974Natur.248...30H,MacGibbon:1991tj}, the total energy injection rate per unit volume, given by $(\mathrm{d}E/\mathrm{d}V \mathrm{d}t)_\mathrm{inj} \propto n_\mathrm{PBH} \dot{M}_\mathrm{PBH}$, scales as $M_\mathrm{PBH}^{-3}$ as the PBH number density follows $n_\mathrm{PBH} \propto M_\mathrm{PBH}^{-1}$. As a result, energy injection decreases with increasing $M_\mathrm{PBH}$, leading to weaker constraints. Here, we note that the constraint on PBH parameters assumes a monochromatic mass distribution. For an extended mass distribution, such as the commonly used log-normal mass function~\cite{Dasgupta:2019cae}, the constraint would extend over a broader mass range, probing more into the asteroid-mass range $( 10^{17}\,\textrm{g}\leq M_{\rm PBH} \leq 10^{22}\,\textrm{g}$).

For comparison, we show previous constraints based on the IGM temperature, from Ref.~\cite{2024arXiv240910617S} by the gray curve in Figure~\ref{fig4}, where we have adopted their strongest reported constraint from the `photoheating' model variations. Our constraints are stronger than these across all mass ranges, with the improvement being more than an order of magnitude for some PBH mass values. As we discussed above, this improvement arises from our inclusion of low-redshift temperature data, where PBH-induced heating via Coulomb interactions is most significant. Additionally, our constraints are also different due to the incorporation of He$\,$\textsc{iii} evolution and a more realistic reionization model \cite{2019MNRAS.485...47P}. 

Relative to other constraints in the literature (not shown in Figure~\ref{fig4}, in order to keep the figure accessible), our constraints are the second strongest in the mass range \( 7 \times 10^{15} \) g to \( 2 \times 10^{17} \) g \cite{Carr:2009jm, Boudaud:2018hqb, DeRocco:2019fjq, Laha:2019ssq,Dasgupta:2019cae, Laha:2020ivk, Ray:2021mxu, Coogan:2020tuf} (a comprehensive comparison with existing bounds is presented in Appendix~\ref{appendix4}). The most stringent limit in this mass window currently comes from measurements of the 511 keV gamma-ray line \cite{2024PhRvD.110l3022L}. While this provides a stronger constraint, it crucially depends on astrophysical modeling of positron propagation in the Galactic magnetic fields, and dark matter density profiles, introducing  significant systematic uncertainties 
(see Ref.~\cite{DeRocco:2019fjq,Laha:2019ssq,Dasgupta:2019cae} for weaker constraints derived from the 511 keV measurement and discussions on how this constraint can relax under various astrophysical uncertainties). In comparison, the Lyman-$\alpha$ forest has been extensively used to derive a wide range of cosmological constraints, demonstrating its precision and robustness as a probe of the IGM. Our IGM temperature constraints, based on well-understood cosmological heating and cooling processes, are independent of local dark matter density and astrophysical backgrounds. Additionally, while the 511 keV gamma-ray signal relies on indirect annihilation signatures, IGM temperature measurements offer a more direct window into PBH-induced heating. Given these distinctions, the Lyman-$\alpha$ forest may provide a complementary and potentially more reliable avenue for constraining PBHs.

Note that, in Figure~\ref{fig4}, we consider PBHs with masses greater than $ \sim 10^{15}$ g, as those below this threshold would have evaporated by now, given their lifetimes are shorter than the age of the Universe~\cite{Page:1976df,Page:1976ki,MacGibbon:1990zk,MacGibbon:2007yq}. 
This sets the lower limit for PBH masses in our analysis. However, see recent studies~\cite{Dvali:2020wft,Thoss:2024hsr,Dvali:2024hsb,Basumatary:2024uwo} for discussions on how quantum-gravity effects might still allow PBHs with $ M_{\rm PBH} \ll 10^{15}$~g to exist today.

\section{Conclusions}

We have self-consistently modeled the temperature and ionization histories of the IGM, incorporating both astrophysical reionization and PBH energy injection. By comparing our predictions with Lyman-$\alpha$ forest temperature measurements, we have placed the second strongest constraints on PBH abundance in the mass range \( 7 \times 10^{15} \) g to \( 2 \times 10^{17} \) g, improving upon previous IGM temperature-based limits by approximately an order of magnitude. This improvement arises from our inclusion of lower-redshift data, where PBH-induced heating via Coulomb interactions is most significant, as well as our extension of the analysis to redshift zero, full incorporation of He$\,$\textsc{iii} evolution, and use of a physically motivated reionization model. We have also slightly expanded the exploration of PBH parameter space to the lower end of the mass range (\( 10^{17} \, \mathrm{g} \) to \( 10^{22} \, \mathrm{g} \)), where no prior constraints were available (see, however, recent proposals~\cite{Ray:2021mxu,Esser:2022owk,Tamta:2024pow,Jung:2019fcs,Gawade:2023gmt,Fedderke:2024wpy} to cover this mass window). While the strongest constraints on PBHs currently come from the 511 keV gamma-ray line, they rely on complex and uncertain astrophysical modeling, including positron propagation in Galactic magnetic fields, and dark matter density profiles. In contrast, our IGM temperature constraints are based on well-understood cosmological heating and cooling processes, independent of local PBH clustering and astrophysical uncertainties. Since PBH energy injection has the strongest impact at low redshifts, more precise future measurements of the IGM temperature at low redshifts (e.g., mock \( T_\mathrm{IGM} \) data at \( z=0.1 \) from \cite{2022MNRAS.515.2188H}) could further improve these constraints. More broadly, this study establishes a framework for leveraging the IGM thermal history to constrain a wide range of exotic energy injection scenarios, providing a powerful probe for testing beyond-Standard Model physics, including various dark matter models. 

\section*{Acknowledgments}
We thank Sulagna Bhattacharya, James Bolton, Fred Davies, Prakash Gaikwad, Martin Haehnelt, Vid Ir\v{s}i\v{c}, Hongwan Liu, and Ewald Puchwein for useful discussions. AR acknowledges support from the National Science Foundation (Grant No.\ PHY-2020275) and to the Heising-Simons Foundation (Grant No.\ 2017-228). GK gratefully acknowledges support by the Max Planck Society via a partner group grant. GK \& BD are also partly supported by the Department of Atomic Energy (Government of India) research project with Project Identification Number RTI 4002.
\bibliographystyle{bibi}
\bibliography{biblio}

\appendix
\section{Standard Heating and Cooling Processes in the IGM} \label{appendix1}
\noindent In this section, we present the key terms governing the temperature and ionization evolution of the IGM, considering only standard astrophysical processes without PBH contributions. These include adiabatic cooling due to Hubble expansion, Compton heating/cooling, atomic processes, and photoheating from reionization.

\subsection{Temperature Evolution}

\noindent The adiabatic cooling term due to cosmic expansion is given by:
\begin{equation}
    \dot{T}_\mathrm{adia} = -2 H T_\mathrm{IGM},
\end{equation}
where $H$ is the redshift-dependent Hubble constant.
The Compton heating/cooling term accounts for energy exchange between free electrons and the CMB:
\begin{equation}
    \dot{T}_\mathrm{C} = \Gamma_C (T_\mathrm{CMB} - T_\mathrm{IGM}),
\end{equation}
where $\Gamma_C$ is the Compton scattering rate, and $T_\mathrm{CMB}$ is the cosmic microwave background temperature.
Atomic cooling processes include collisional ionization, recombination, collisional excitation, and Bremsstrahlung cooling:
\begin{align}
    \dot{T}_\mathrm{atom} &= \frac{2}{3 (1 + \mathcal{F}_\mathrm{He} + x_e) n_\mathrm{H}}  \nonumber \\
    & \quad \times \sum_{X} \left(\mathcal{H}^{\mathrm{ion}}_{eX} + \mathcal{H}_{X}^{\mathrm{rec}} + \mathcal{H}_{eX}^{\mathrm{exc}} + \mathcal{H}^{\mathrm{br}} \right).
\end{align}
The summation runs over relevant species: $X = \mathrm{H}\,\textsc{i}, \mathrm{He}\,\textsc{i}, \mathrm{He}\,\textsc{ii}$ for collisional ionization and excitation, and $X = \mathrm{H}\,\textsc{ii}, \mathrm{He}\,\textsc{ii}, \mathrm{He}\,\textsc{iii}$ for recombination and Bremsstrahlung cooling. The cooling rates $\mathcal{H}$ are computed following Appendix B4 of Ref.~\cite{2007MNRAS.374..493B}.
Photoheating from reionization is given by:
\begin{align}
    \dot{T}_\mathrm{reion} &= \frac{2}{3 (1 + \mathcal{F}_\mathrm{He} + x_e) n_\mathrm{H}} \nonumber \\ 
    &\quad \times n_\mathrm{H} (x_\mathrm{HI} \varepsilon_\mathrm{HI} + x_\mathrm{HeI} \varepsilon_\mathrm{HeI} + x_\mathrm{HeII} \varepsilon_\mathrm{HeII}).
\end{align}
The photoheating rates $\varepsilon$ are taken from Table D1 of Ref.~\cite{2019MNRAS.485...47P}.

\subsection{Ionization Evolution}

\noindent The evolution of hydrogen and helium ionization fractions is governed by atomic and photoionization processes. For hydrogen ionization:
\begin{equation}
    \dot{x}_{\mathrm{HII}}^{\mathrm{atom}} = x_{\mathrm{HI}}  n_e \Gamma_{e \mathrm{HI}}^{\mathrm{ion}}  - x_{\mathrm{HII}} n_e \alpha_{A,\mathrm{HI}},
\end{equation}
where $\Gamma_{e \mathrm{HI}}^{\mathrm{ion}}$ is the collisional ionization rate and $\alpha_{A,\mathrm{HI}}$ is the Case-A recombination coefficient, both taken from Appendix B4 of Ref.~\cite{2007MNRAS.374..493B}.
The contribution from reionization is:
\begin{equation}
    \dot{x}_{\mathrm{HII}}^{\mathrm{reion}} = x_{\mathrm{HI}}  \Gamma_{\gamma \mathrm{HI}},
\end{equation}
where $\Gamma_{\gamma \mathrm{HI}}$ is the photoionization rate of H$\,$\textsc{i}, taken from Table D1 of Ref.~\cite{2019MNRAS.485...47P}.
For He$\,$\textsc{ii} ionization:
\begin{align}
    \dot{x}_{\mathrm{HeII}}^{\mathrm{atom}} &= x_{\mathrm{HeI}}  n_e \Gamma_{e \mathrm{HeI}}^{\mathrm{ion}}  + x_{\mathrm{HeIII}} n_e \alpha_{A,\mathrm{HeIII}} \nonumber \\
    &\quad - x_{\mathrm{HeII}} \left( n_e \Gamma_{e \mathrm{HeII}}^{\mathrm{ion}} + n_e \alpha_{A,\mathrm{HeII}} \right),
\end{align}
where He$\,$\textsc{ii} increases due to the ionization of He$\,$\textsc{i} and recombination of He$\,$\textsc{iii}, and decreases due to the ionization of He$\,$\textsc{ii} and recombination of He$\,$\textsc{ii}. The collisional ionization rates and Case-A recombination coefficients are taken from Appendix B4 of Ref.~\cite{2007MNRAS.374..493B}.
The reionization contribution to He$\,$\textsc{ii} is:
\begin{equation}
    \dot{x}_{\mathrm{HeII}}^{\mathrm{reion}} = x_{\mathrm{HeI}}  \Gamma_{\gamma \mathrm{HeI}}^{\mathrm{ion}}  - x_{\mathrm{HeII}}  \Gamma_{\gamma \mathrm{HeII}}^{\mathrm{ion}},
\end{equation}
where the photoionization rates are taken from Table D1 of Ref.~\cite{2019MNRAS.485...47P}.

\section{Coulomb Heating and Energy Deposition} \label{appendix2}

\noindent High-energy electrons and positrons injected by PBHs deposit energy into the IGM through various processes. After reionization, the dominant energy deposition channel is Coulomb heating, where energetic electrons transfer their kinetic energy to background free electrons, increasing the IGM temperature. Additionally, injected energy is also distributed into collisional ionization, collisional excitation, and inverse Compton scattering. In this work, we follow the methodology of Ref.~\cite{2020PhRvD.101b3530L} to model the heating due to PBH-injected electrons and positrons. This section specifically focuses on how we recomputed $f_{\rm heat}(z,{\textbf x})$ (as defined in Equation~\ref{equ7} in the main text) that encodes the fraction of injected energy which converts to heat.

\subsection{Coulomb Heating in the IGM}

\noindent After reionization, the abundance of free electrons makes Coulomb interactions the most efficient energy deposition channel. The energy loss rate of an electron/positron of energy $E_n$ due to Coulomb interactions is given by:
\begin{equation}
\frac{\mathrm{d} E_\mathrm{heat}}{\mathrm{d} t} = \frac{e^4  \ln{\Lambda}}{4 \pi \epsilon_0^2 m_e v}n_e, \label{S1}
\end{equation}
where the free electron number density
\begin{equation}
n_e = n_\mathrm{H} (x_\mathrm{HII} + x_\mathrm{HeII} + 2x_\mathrm{HeIII})
\end{equation}
is determined by the ionization state ($\textbf x$) of the IGM, with $e$ being the elementary charge, \( \ln{\Lambda} \) the Coulomb logarithm accounting for the range of impact parameters in collisions, \( \epsilon_0 \) is the permittivity of free space, \( m_e \) is the electron mass, and \( v \) is the velocity of the injected electron or positron. Since Coulomb interactions efficiently thermalize injected electrons, most of their energy is converted into heat. The fraction of injected energy deposited as heat, represented by \( f_{\text{heat}}(z, \textbf{x}) \), varies with redshift \( z \) and is influenced by the ionization state of the IGM, determining how effectively PBH-injected energy impacts IGM heating.

%\subsection{Energy Redistribution in Electron Spectra}  

Over a small time step \( \Delta t \), corresponding to the time step used in the temperature and ionization evolution equations, the energy lost by an electron due to Coulomb interactions is given by:  

\begin{equation}
\Delta E_\mathrm{heat} = \Delta t \frac{\mathrm{d} E_\mathrm{heat}}{\mathrm{d}t}.
\end{equation}
Initially, an electron in energy bin \( E_n \) has a probability \( P_n = 1 \), meaning it is fully present in that energy state. As it loses energy, it transitions to a lower energy state, redistributing its energy. PBH-injected electron or positron has very high energy, its velocity $v$ is large. Additionally, by choosing a sufficiently small time step $\Delta t$, the energy loss $\Delta E_\mathrm{heat}$ can be made small. As a result, in a single time step, an electron initially in energy $E_n$ will transition only to the immediate lower bin $E_{{n-1}}$ after losing energy. This redistribution follows energy conservation:  
\begin{equation}
P_{n-1} E_{n-1} + P_n E_n = E_n - \Delta E_\mathrm{heat}, \label{S2}
\end{equation}
where \( P_n \) is the probability that the electron remains in its current energy bin, and \( P_{n-1} \) is the probability that it moves to the adjacent lower-energy bin. Since the electron must exist in one of these states, we have:  
\begin{equation}
P_{n-1} + P_n = 1. \label{S3}
\end{equation}  
Solving for \( P_n \) and \( P_{n-1} \), we obtain:  
\begin{equation}
P_n = 1 - \frac{\Delta E_\mathrm{heat}}{E_n - E_{n-1}}, \quad
P_{n-1} = \frac{\Delta E_\mathrm{heat}}{E_{n} - E_{n-1}}. \label{S4}
\end{equation}  
This ensures that energy loss is accurately tracked as electrons cascade down in energy, preserving energy conservation.  
Based on the above calculation, the redistribution of electrons due to energy loss can be expressed as a matrix equation:
\begin{equation}
    \mathbf{\frac{\mathrm{d}N_\mathrm{sec}}{\mathrm{d}E}} = \mathbf{M} \cdot \mathbf{\frac{\mathrm{d}N}{\mathrm{d}E}}, \label{eng_matrix}
\end{equation}
where:
\begin{itemize}
    \item[-] \( \mathbf{\frac{\mathrm{d}N_\mathrm{sec}}{\mathrm{d}E}} \) is the secondary electron spectrum after energy loss.
    \item[-] \( \mathbf{\frac{\mathrm{d}N}{\mathrm{d}E}} \) is the electron spectrum before energy loss.
    \item[-] \( \mathbf{M} \) is the transition matrix describing how electrons move between energy bins.
\end{itemize}
The transition matrix \( \mathbf{M} \) is given by:
\begin{equation}
    \mathbf{M} =
    \begin{bmatrix}
        P_0 & P_{0,1} & 0 & 0 & \dots & 0 \\
        0 & P_1 & P_{1,2} & 0 & \dots & 0 \\
        0 & 0 & P_2 & P_{2,3} & \dots & 0 \\
        0 & 0 & 0 & P_3 & \dots & 0 \\
        \vdots & \vdots & \vdots & \vdots & \ddots & P_{N-2,N-1} \\
        0 & 0 & 0 & 0 & 0 & P_{N-1}
    \end{bmatrix}, \label{matrix}
\end{equation}
where:
\begin{align}
    P_n &= 1 - \frac{\Delta E_{\mathrm{heat}}}{E_n - E_{n-1}}, \quad \text{(diagonal elements)}, \\
    P_{n-1,n} &= \frac{\Delta E_\mathrm{heat}}{E_n - E_{n-1}}, \quad \text{(upper-diagonal elements)}.
\end{align}
Here, we consider \( N \) energy bins labeled \( 0, 1, 2, 3, ..., N-1 \).  
All elements below the diagonal are zero, ensuring that electrons can only lose energy.

\subsection{Recursive Energy Deposition Framework}  

\noindent Let \( R_\mathrm{heat}(E_n) \) represent the total energy deposited into Coulomb heating, accounting for the cumulative energy transferred from the injected electron until its energy is fully dissipated. Over the short time interval \( \Delta t \), the injected electron undergoes various cooling processes, including Coulomb interactions, collisional ionization, collisional excitation, and inverse Compton scattering. These interactions produce a secondary electron spectrum \( \mathrm{d}\textbf{N}_\mathrm{sec}/\mathrm{d}\textbf{E} \) (see Equations~\ref{eng_matrix} and \ref{matrix} for the case considering only Coulomb heating, and Ref.~\cite{2009PhRvD..80d3526S} for cross sections and secondary spectra associated with other processes). A fraction of the injected electron’s energy, \( \Delta E_\mathrm{heat} \) (given by Equation~\ref{S1}), is promptly deposited as heat within the same time interval \( \Delta t \). The remaining energy is carried by the secondary electron, which subsequently loses energy through Coulomb interactions, described by \( R_\mathrm{heat}(E) \), where \( E \) represents the energy of the secondary electron. This process leads to the following recursive relation:
\begin{equation}
    R_\mathrm{heat} (E_n) = \int \mathrm{d}E \, R_\mathrm{heat} (E) \frac{\mathrm{d}\textbf{N}_\mathrm{sec}}{\mathrm{d}\textbf{E}} + \Delta E_\mathrm{heat}. \label{S5}
\end{equation}  
This equation describes the total energy deposited into Coulomb heating, consisting of:
\begin{enumerate}
    \item The energy directly deposited by the injected electron, \( \Delta E_\mathrm{heat}\) within the time interval $\Delta t$.
    \item The energy deposited by the secondary electron as it further loses energy through Coulomb interactions, represented by \( R_\mathrm{heat}(E) \).   
\end{enumerate}
To obtain \( R_\mathrm{heat}(E_n) \), we numerically solve Equation~\ref{S5}, using the known values of \( \Delta E_\mathrm{heat}\) and \( \mathrm{d}\textbf{N}_\mathrm{sec}/\mathrm{d}\textbf{E} \). This iterative approach ensures self-consistency while conserving both energy and particle number. With the obtained \( R_\mathrm{heat}(E_n) \), we can now compute \( f_{\text{heat}}(z, \textbf{x}) \) using Equation~\ref{equ8} from the main text.

\begin{figure}
    \centering
    \includegraphics*[width=0.5\textwidth]{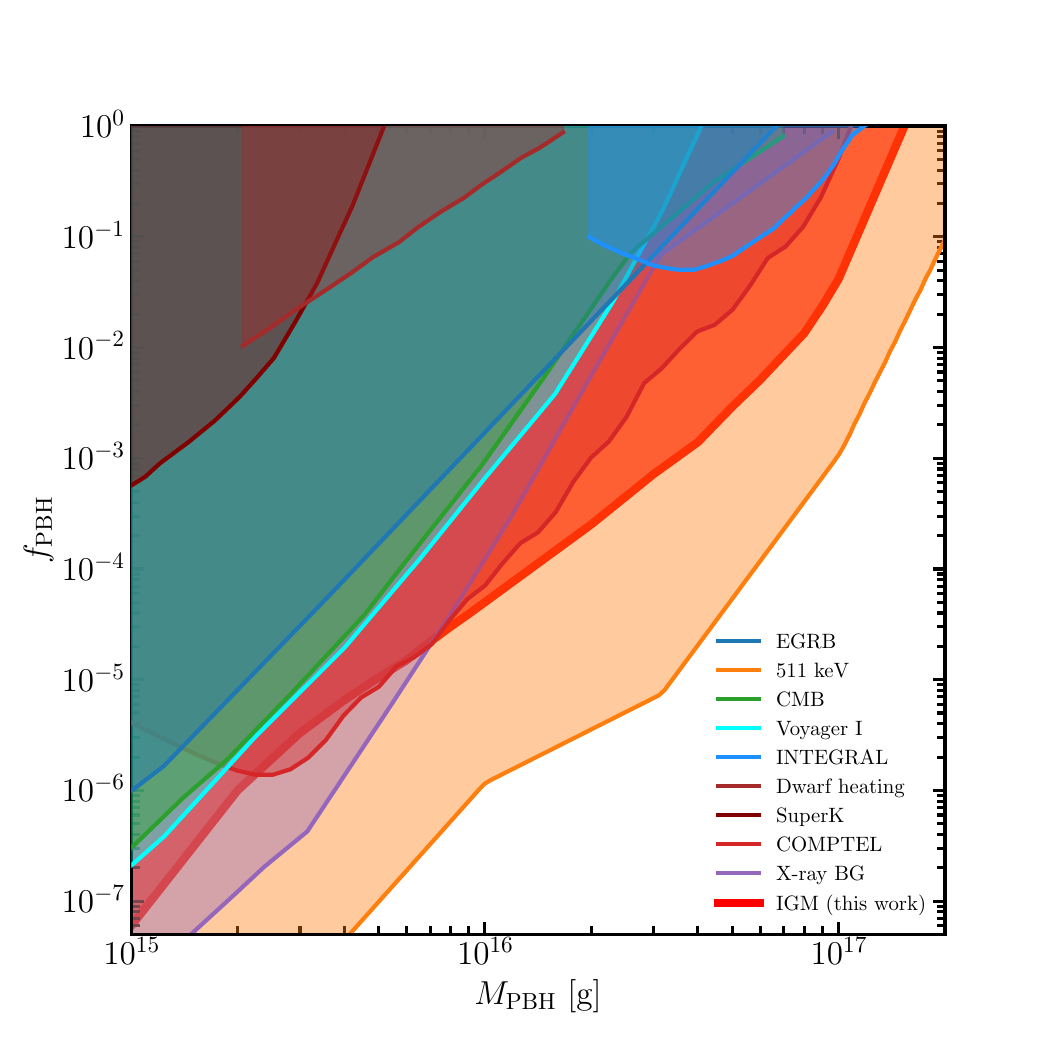}
    \caption{{Constraints on the PBH fraction $f_\mathrm{PBH}$ as a function of PBH mass $M_\mathrm{PBH}$. Our new constraint (red solid line) is shown alongside existing bounds from: extragalactic gamma rays \cite{Carr:2009jm} in blue, the 511 keV gamma-ray line \cite{2024PhRvD.110l3022L} in orange, modifications of the CMB spectrum \cite{Clark:2016nst} in green, electron/positron measurements from Voyager 1 \cite{Boudaud:2018hqb} in cyan, the MeV background \cite{Laha:2020ivk} in dodger blue, heating effects in the dwarf galaxy Leo T \cite{Laha:2020vhg} in brown, neutrino evaporation limits from Super-Kamiokande \cite{Dasgupta:2019cae} in maroon, COMPTEL GC observations \cite{Coogan:2020tuf} in red, and the X-ray background \cite{2024PhRvD.110l3022L} in purple. Some of these constraints, including the 511 keV one (orange), depends on assumed dark matter density profile and can therefore be relaxed under different profile choices, whereas our constraint is based on the cosmological dark matter density and is independent of such uncertainties.}}
    \label{fig5}
\end{figure}

\section{Table of Constraints on \( f_{\mathrm{PBH}} \)} \label{appendix3}
\noindent We present the numerical values of the upper limits on \( f_{\mathrm{PBH}} \) derived from our analysis. The table below (see Table~\ref{tab:constraints}) lists the constraints for different PBH masses, ranging from \( 1 \times 10^{15} \) g to \( 2 \times 10^{17} \) g. 
\begin{table}[H]
    \centering
    \begin{tabular}{c @{\hskip 20pt} c}
        \hline
        \( M_{\mathrm{PBH}} \) [g] &  \( f_\mathrm{PBH}\) \\ 
       \hline \vspace{-0.25 cm} \\ 
        \( 1 \times 10^{15} \)  & \( 5.9979 \times 10^{-8} \) \\
        \( 2 \times 10^{15} \)  & \( 1.0034 \times 10^{-6} \) \\
        \( 3 \times 10^{15} \)  & \( 3.3068 \times 10^{-6} \) \\
        \( 4 \times 10^{15} \)  & \( 6.5185 \times 10^{-6} \) \\
        \( 5 \times 10^{15} \)  & \( 1.0445 \times 10^{-5} \) \\
        \( 6 \times 10^{15} \)  & \( 1.4972 \times 10^{-5} \) \\
        \( 7 \times 10^{15} \)  & \( 2.1965 \times 10^{-5} \) \\
        \( 8 \times 10^{15} \)  & \( 3.0132 \times 10^{-5} \) \\
        \( 9 \times 10^{15} \)  & \( 3.9209 \times 10^{-5} \) \\
        \( 1 \times 10^{16} \)  & \( 5.0130 \times 10^{-5} \) \\
        \( 2 \times 10^{16} \)  & \( 2.5286 \times 10^{-4} \) \\
        \( 3 \times 10^{16} \)  & \( 7.1961 \times 10^{-4} \) \\
        \( 4 \times 10^{16} \)  & \( 1.3954 \times 10^{-3} \) \\
        \( 5 \times 10^{16} \)  & \( 2.9074 \times 10^{-3} \) \\
        \( 6 \times 10^{16} \)  & \( 5.0939 \times 10^{-3} \) \\
        \( 7 \times 10^{16} \)  & \( 8.5776 \times 10^{-3} \) \\
        \( 8 \times 10^{16} \)  & \( 1.3460 \times 10^{-2} \) \\
        \( 9 \times 10^{16} \)  & \( 2.3649 \times 10^{-2} \) \\
        \( 1 \times 10^{17} \)  & \( 4.1475 \times 10^{-2} \) \\
        \( 2 \times 10^{17} \)  & \( 7.0747 \) \\
        \hline
    \end{tabular}
    \caption{Constraints on \( f_\mathrm{PBH} \) as a function of PBH mass, obtained from IGM temperature measurements. These constraints are derived using a $\chi^2$ difference test at the 95\% confidence level. The resulting exclusion limits are shown as the red curve in Figure~\ref{fig4}.}
    \label{tab:constraints}
\end{table}

\section{Comparison with Existing Constraints on \( f_{\mathrm{PBH}} \)} \label{appendix4}
\noindent{In this appendix, we present a comprehensive comparison of our newly derived constraint on \( f_{\mathrm{PBH}} \) with various existing bounds reported in the literature. Figure~\ref{fig5} displays these constraints as a function of \( M_{\mathrm{PBH}} \) for a monochromatic mass function of PBHs.}

\end{document}